\documentclass[openacc]{rstransa}

\begin{document}

\title{The dynamics of 3-minute wavefronts and their relation to sunspot magnetic fields}

\author{
Robert Sych$^{1}$, David B. Jess$^{2}$ and Jiangtao Su$^{3,4}$}

\address{$^{1}$ Institute of Solar-Terrestrial Physics SB RAS, Lermontov st. 126a, Irkutsk 664033, Russia\\
$^{2}$ Astrophysics Research Centre, School of Mathematics and Physics, Queen$'$s University Belfast, Belfast, BT7 1NN, UK \\
$^{3}$ Key Laboratory of Solar Activity, National Astronomical Observatories, Chinese Academy of Science, Beijing 100101, China\\
$^{4}$ School of Astronomy and Space Sciences, University of Chinese Academy of Sciences, 19A Yuquan Road, Beijing 100049, China}

\subject{solar physics, wave propagation, mathematical modelling}


\corres{Robert Sych\\
\email{sych@iszf.irk.ru}}

\begin{abstract}
We present a study of wave processes occurring in solar active region NOAA~11131 on 2010 December 10, captured by the Solar Dynamics Observatory in the 1600{\AA}, 304{\AA}, and 171{\AA} channels. For spectral analysis we employed pixelised wavelet filtering together with a developed digital technique based on empirical mode decomposition. We studied the $\sim$3-minute wave dynamics to obtain relationships with the magnetic structuring of the underlying sunspot. We found that during development of wave trains the motion path occurred along a preferential direction, and that the broadband wavefronts can be represented as a set of separate narrowband oscillation sources. These sources become visible as the waves pass through the umbral inhomogeneities caused by the differing magnetic field inclination angles. We found the spatial and frequency fragmentation of wavefronts, and deduced that the combination of narrowband spherical and linear parts of the wavefronts provide the observed spirality. Maps of the magnetic field inclination angles confirm this assumption. We detect the activation of umbral structures as the increasing of oscillations in the sources along the front ridge. Their temporal dynamics are associated with the occurrence of umbral flashes. Spatial localisation of the sources is stable over time and depends on the oscillation period. We propose that these sources are the result of wave paths along the loops extending outwards from the magnetic bundles of the umbra. 
\end{abstract}

\maketitle

\section{Introduction}
Waves that are generated at the sub-photospheric level propagate through natural magnetic waveguides in the form of sunspots and pores, with their oscillation period being modified by the local cutoff frequency. It has been shown \cite{1} that oscillations with frequencies lower than the magnetoacoustic cutoff frequency quickly attenuate. This leads to the channelling of oscillations into the upper layers of the solar atmosphere \cite{2,3}. The main factor affecting the specific cutoff frequency is the inclination angle of the magnetic field lines that guide the wave motion and temperature. In \cite{53} it was shown that the temperature changes have a smaller effect on the wave frequency distribution across the sunspot than the magnetic field inclination angles. The propagation speed of disturbances in sunspots indicates the presence of slow magnetoacoustic waves \cite{4}.  Oscillation sources with periods less than $\sim$3 minutes are localised in the umbra and reduce in physical size as the period decreases \cite{5,6,7}. In the central part of the umbra, the footpoints of thin magnetic tubes appear in the form of oscillating cells and dots \cite{8}.  One of the main physical mechanisms that determines the oscillatory power in the umbral core is related to the resonant cavity trapped between the photosphere and chromosphere \cite{9, Jess2020}. Increasing angles of magnetic field inclination towards the sunspot periphery result in direct changes to the associated cutoff frequency \cite{1}.

Sunspot oscillations are regularly visible as travelling waves that propagate radially outwards from the centre of the umbra. During propagation, the shape of the wavefronts can transform and interchange between spherical and spiral structures. The spiral wave patterns are visible at all altitudes above the sunspot, and it was hypothesised that the local dependence of the magnetic field inclination angle on the polar angle may play a role in the wavefront signatures \cite{5}. It has been suggested that the single-spiral structure of  the wavefronts may be caused by reflections at an umbral light bridge \cite{10}, while the multi-spiral structuring may be associated with swirling magnetic fields beneath the umbral photosphere. It has been shown that wavefronts initially move in a counter-clockwise direction in the interior of the umbra, before gradually developing into radial running penumbral waves (RPWs) \cite{11}. The propagation of horizontal waves is related to the direction of magnetic fields, essentially dependent on the magnetic vortex created across the umbral space and with geometric height. It is assumed that the development of wavefronts can be caused by a thin magnetic structure in the umbra, including umbral dots and light bridges, where the cutoff frequency may be inherently different. It has been assumed that the waveguides of running waves are twisted into local spiral structures \cite{12}. A model for the appearance of spiral wavefronts was proposed in relation to their propagation in radial magnetic fields \cite{13}. The authors concluded that rotating structures can occur in a superposition of non-zero-$m$ azimuthal slow modes. Another model for the appearance of spiral wavefronts has been proposed, concluding that the rotating oscillatory power was an $m=1$ slow kink mode superimposed on top of the ubiquitous $p$-mode oscillations \cite{14}. Period drifts, on the order of $90 - 240$~seconds, have been observed during the development stage of quasi-spiral wavefronts \cite{15}. The beginning and end of the period drifts coincide with the evolutionary dynamics of low-frequency wave trains.  

In the umbra, along with running waves, there are also so-called umbral flashes (UFs; \cite{16,17}). It is assumed that they are caused by upwardly propagating magnetoacoustic waves that steepen with altitude and turn into shock fronts \cite{18}. As such, propagating waves inherent to sunspots are not stationary, and evolve in power through both time and space \cite{19, Houston2018}. This leads to low-frequency modulation of the wave signatures as the wave trains superimpose with peak UF non-linearities to form maximum power enhancements. Only recently have the first images of thin wave processes in horizontal magnetic structures associated with UFs been obtained \cite{20}. It was shown that there are two types of UFs, linked to the `background' and `local' environments. The first is related to the strengthening of individual parts of the wavefronts, corresponding to weak and diffuse emission on the crest of propagating fronts. Local UFs are, by definition, localised near the footpoints of thin magnetic waveguides, anchored in the umbra. Recently, it was shown that in the chromosphere the umbral oscillations are a collection of large independent oscillations located in isolated sources with small angular size \cite{21}, resulting in their spatial structure changing from cellular pockets at the umbral centre through to filamentary features at the outer edge of the sunspot periphery. Each narrow harmonic of the corresponding spectrum relates to a source that is connected to a separate magnetic flux tube anchored in the umbra.
		
The magnetic field of sunspots is predominantly vertical in nature near the central part of umbra. This assumption has been confirmed by polarimetric observations \cite{22,23}. In a number of publications, the fine fragmentation of umbral magnetic fields was found in regions with a strong underlying magnetic field \cite{24,25}. Based on optical data, it has been shown that 5-min oscillations are mainly concentrated in isolated magnetic field conduits (e.g., pores) at the boundaries of the magnetic structures \cite{26}. Observations with high spatial resolution have shown the presence of very small horizontal jet-like spatial structures in umbrae with a size less than 0.1~Mm \cite{27}, with their positions coinciding with the magnetic waveguides. A similar distribution has been modelled for the general case of a magnetic atmosphere \cite{28}, concluding that the modelled properties of these fine-scale details are similar to those previously observed \cite{29}. Long-lived horizontal structures have been observed in Ca~{\sc{ii}} data to exist in sunspot umbrae with a size $\sim 0.15''$, which were illuminated by developing UFs. It is hypothesised that such Ca~{\sc{ii}} thread-like structures are a common phenomenon within the spatial structure of sunspot umbrae. At the photosphere level, there are fine-scale structures with opposite magnetic polarity in the innermost regions of the penumbra \cite{31}, with similar structures also seen in the chromosphere. The chromospheric horizontal structures \cite{29} are similar in properties to dynamic fibrils \cite{32}, but with a smaller angular size. 

All of the fine-scale structures now readily visible inside sunspot umbrae allow us to revise the general idea regarding the ``verticality'' of sunspot magnetic fields. The existence of thin horizontal magnetic loops subtending the umbral core, which are embedded within ubiquitous propagating waves, provides us with an ideal testbed for examining the composition of a sunspot atmosphere.

In our work, we analyse the relationship between the spatial distribution of propagating wavefronts and study their interaction with the embedded magnetic structures with sizes ranging from $\sim 0.6''$ up to $\sim 3''$ (435 -- 2175~km). Using pixelised wavelet filtering techniques \cite{33}, we reconstruct two-dimensional maps of the magnetic field inclination based on the study of the wave cutoff frequency distribution. This allows us to interpret the spatial behaviour of spiral wavefronts during radial propagation, as well as to understand the occurrence of background UFs and the appearance of the $\sim$3~minute frequency drift. We believe that understanding the field structure in sunspots is important for both building a model of energy transfer from the sub-photospheric layers, and for understanding the possible heating of the corona by wave processes.
	
The structure of the article is as follows: in Section~1 we give an introduction on the topic and present the objectives of the research; in Section~2 we describe the observations and data processing; Section~3 shows the obtained results and discusses the interpretations associated with the spatial, frequency, and temporal dynamics of wavefronts in our sunspot. The magnetic structure of the umbra obtained from our study is also considered; and in Section~4 we present the conclusions of our work.


\section{Observations and Data processing}
 We used image cubes of sunspot active region NOAA~11131 obtained in the ultraviolet (UV) and extreme~UV (EUV) bands using the Atmospheric Imaging Assembly (AIA; \cite{34}) aboard the Solar Dynamics Observatory (SDO). The observations were acquired on 2010 December 10 from 00:00 -- 01:00~UT in the temperature channels 1600{\AA} (UV), 304{\AA} (EUV), and 171{\AA} (EUV). The SDO/AIA online resource\footnote{NASA/SDO data archive: \url{http://www.lmsal.com/get_aia_data/}} was used to obtain Level~1 images of the Sun in each of the different wavelengths, with a subsequent spatial sampling of $0{\,}.{\!\!}{''}6$ per pixel. To compare data between different channels, a time resolution of 24~seconds was selected. The duration of the observation was one hour, which allowed us to study oscillations in the range of periods from $\sim$1 -- 20~minutes. An umbral region with an angular size of $18'' \times 18''$ was selected, with final calibration of the images (to Level~1.5) performed to remove differential rotation\footnote{performed using the algorithms present in the SolarSoft library: \url{https://sohowww.nascom.nasa.gov/solarsoft/}}.
 
\subsection{Method of Pixelised Mode Decomposition}
To search for propagating waves in the sunspot and recover their temporal dynamics, we employed a pixelised wavelet filtering (PWF) technique \cite{33}, which is widely used for studying oscillations in coronal loops \cite{47,48} and determining the fine spatial structure of wave sources in sunspots \cite{8}. This technique is a generalisation of the wavelet transform techniques applied to image cubes. For all pixels studied we constructed the wavelet spectra using the Morlet mother function. The obtained data is transformed into four-dimensional narrowband maps  (two-dimensional spatial coordinates, plus frequency and time dimensions) of the amplitudes, power, and phase distributions of the oscillation sources. We used a frequency sub-band and/or separated harmonics for calculation of the narrowband oscillation maps in order to track their spatial dynamics with time. The obtained dynamics are visualised by movies created in an {\sc{avi}} format. Selecting a one-dimensional slice through the narrowband map allowed us to obtain time-distance plots or the time dependency of intensities along selected directions. This give us the possibility to identify oscillation modes, separate standing and running waves, determine the spatial morphology of narrowband wave sources, and study their temporal dynamics.

In addition, we have developed a digital technique for spectral processing of wavefronts, allowing the Pixelised Mode Decomposition (PMD) of signals. This digital technique is based on Empirical Mode Decomposition (EMD) algorithms \cite{35} which are often used in solar physics \cite{58,59,60}. These algorithms adaptively and locally decomposes any non-stationary time series into a sum of Intrinsic Mode Functions (IMFs) with choice of the stopping criterium and the parameters for the selection of the maxima and minima, and does not require any a-priori defined basis system. The developed technique has made it possible to search for waves in the sunspot region and track their corresponding directions of propagation.

The PMD technique is similar to the previously developed pixelised wavelet filtering technique \cite{33}. The main difference between the PWF and PMD techniques is that the PMD algorithms do not use convolution of the signal with a predefined mother function \cite{49}. Instead, we employ the IMF components, which are obtained by cyclically calculating all signal extremes and subtracting the average value. This allows us to obtain information that does not depend on the choice of the spectral mother function and its associated parameters. The resulting set of IMF components includes all frequencies present in the original signal. Each mother function in the PWF technique has its own frequency dimension, which during convolution gives frequency broadening and the blurring of oscillation sources. On the other hand, the PMD technique allows us to work only with instantaneous responses at discrete frequencies represented in the true signal. This allows us to obtain a single set of oscillating sources. The frequency band of the signal will be significantly narrower when compared to the case of PWF analysis. This allows us to increase the frequency and spatial resolution of our technique.

Digital analyses of images is divided into several stages. First, the images were co-aligned. Next, the PMD intensity analysis was performed for each pixel of the image with decomposition into its IMF components in the range of periods of interest (e.g. $\sim$3~minute). To obtain the distribution of periods, we constructed histograms of the separation between the obtained IMF extrema. The obtained dependences of the average periods from power of all IMF components for each pixel allowed us to obtain the period distribution of the spectrum depending upon the level of the oscillations.

\begin{figure}[!t]
\vspace*{-7pt}
\centering\includegraphics[width=13cm]{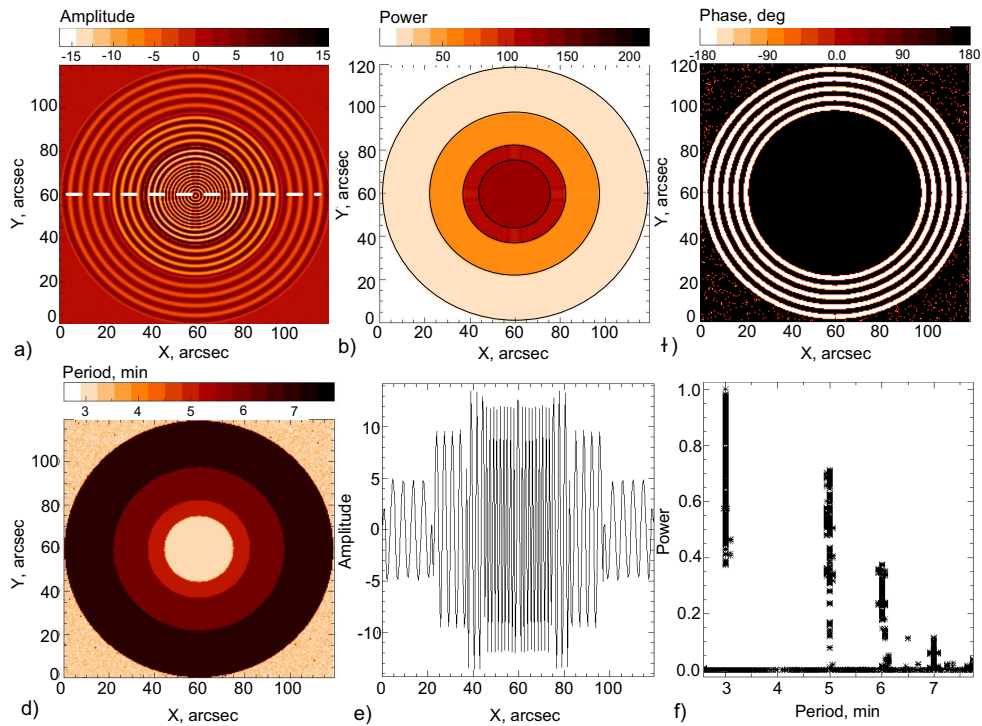}
\caption{Application of the pixelised mode decomposition for artificial test images consisting of spherical running waves. Panel (a) shows a snapshot of the broadband amplitude distribution, while panel (c) shows a snapshot of the phase distribution for a period of 7~minutes. Panels (b) and (d) display the corresponding power and peak period maps, respectively, of the oscillations. Amplitude fluctuations for the central part of the images is shown in the panel (e), where the horizontal dashed line in panel (a) shows the scanning direction. The oscillation spectrum is shown in the panel (f). Oscillation periods are in minutes and spatial coordinates in arcseconds.}
\label{1}
\end{figure}

The resulting histogram is a spectrum that gives us the average and maximum periods present in each pixel. We used this information to select the necessary periodic bands, with signal recovery performed by summing the individual IMF components present within the chosen bandwidth. By repeating this algorithm for all imaged pixels, we replaced the original lightcurves with the filtered ones. For each narrowband signal, the variance was calculated and a power map was prepared. The combination of all periods across all IMF components gives us a final power spectrum. The narrowband reconstructed signals provides us with instantaneous wave amplitude and phases across the duration of the observing sequence. We prepared narrowband oscillation sources and tracked their movement over time, as well as studied the changes in their shape.

\subsection{Model testing}
We tested the PMD technique using a datacube of artificial images created to represent a series of radially diverging spherical waves with amplitudes and periods consistent with those found in sunspots \cite{2013ApJ...779..168J}. Noise was added to the oscillations with 3, 5, 6, and 7 minute periods, each of which included variable amplitudes. The duration of the artificial time series is 100~minutes, with a sample size of $120''\times 120''$. Figure~{\ref{1}} shows the testing results.

\begin{figure}[!t]
\vspace*{-7pt}
\centering\includegraphics[width=13cm]{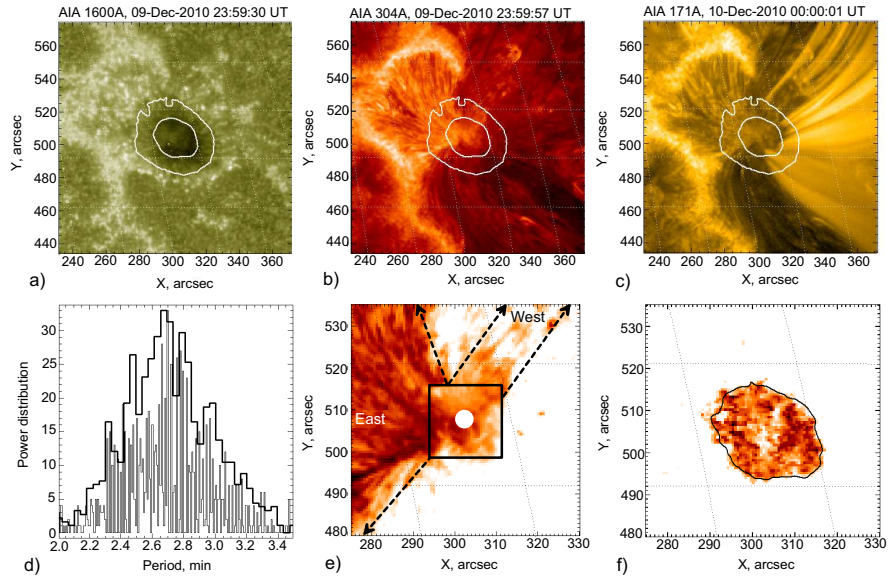}
\caption{Snapshots of the sunspot active region NOAA 11131 obtained in the SDO/AIA 1600{\AA} (a), 304{\AA} (b), and 171{\AA} (c)  channels. The white contours shows the umbra and penumbra boundary in optic. Panel (d) depicts the power distribution of the umbral oscillations. 
The distributions with different bin size are shown by thin and thick lines. Panel (e) displays the intensity variance of the SDO/AIA 304{\AA} observations. The black square shows the position of an $18'' \times 18''$ region of interest, while the white circle shows the location of a pulsating source. The arrows shows the prevailing sectors of wavefronts propagation in Eastern and Western direction. Panel (f) reveals the spatial distribution of the $\sim$3-minute broadband sources in umbra. The panels (e) and (f) displayed in a reverse colour table for visual clarity.}
\label{2}
\end{figure}

The obtained snapshots of the broadband amplitude map (Fig.~\ref{1}a) and the phase map with a period of 7 minutes (Fig.~\ref{1}c) show repeated waves propagating radially from the centre. In a time series based on the narrowband filtered images, running wavefronts with different amplitudes, periods, and phases are observed. The location and shape of the oscillation sources are shown in (Fig.~\ref{1}b) as a power map. Oscillations with a 3-minute period are located in the central part (i.e., the modelled ``umbra''). The 5-minute periodicity occupies an area on the ``umbra/penumbra'' border as a narrow ring. Sources of low-frequency oscillations with periods of 6 and 7 minutes also demonstrate ring shapes with increased angular size. The colour map of the period distribution is shown in (Fig.~\ref{1}d), where four regions with specified periodicities are visible. The intensity profile of instantaneous amplitudes with different periods (Fig.~\ref{1}e) along the central part of the test image (Fig.~\ref{1}a, broken line) coincides with the power distribution shown in the spectrum (Fig.~\ref{1}f).  The obtained results reveal that initial parameters of model oscillations, in terms of their source positions, spatial shape, power, and periods are recovered well by the PMD techniques. Therefore, we are satisfied that this approach can be applied directly to solar observations.

\section{Results and Discussion}

\subsection{Temporal dynamics of the 3-minute broadband wavefronts}

\begin{figure}[!t]
\vspace*{-7pt}
\centering\includegraphics[width=13cm]{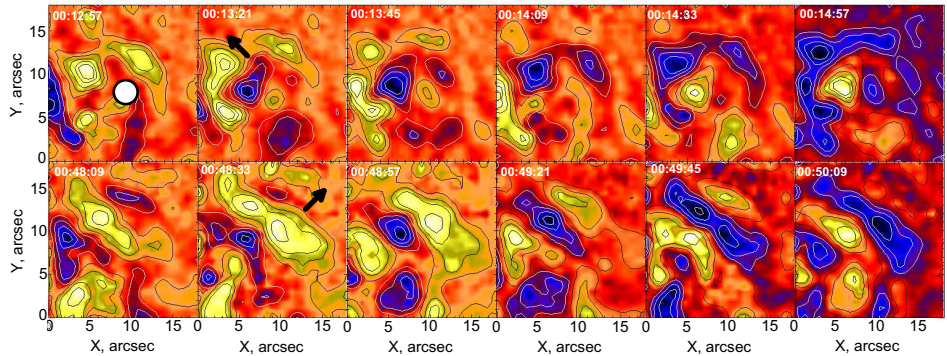}
\caption{Snapshots of the $\sim$3-minute broadband wavefronts propagating in the SDO/AIA 304{\AA} channel from 00:12:57 -- 00:14:57~UT (upper panels) and 00:48:09 -- 00:50:09~UT (lower panels) with a 24 second cadence. Different colours associated with different half-period parts of wavefronts. The arrows show the direction of the wave movements. The white circle shows the localization of the pulsating source. Image brightness is represented on a logarithmic scale.}
\label{3}
\end{figure}

Figure~\ref{2} shows snapshots of sunspot active region NOAA 11131 obtained during the one hour observational period in the 1600{\AA} (panel a), 304{\AA} (panel b), and 171{\AA} (panel c) wavelengths. The white contours shows the umbra and penumbra boundary. We selected a region of interest (ROI) as a square in (Fig.~\ref{2}e) with an angular size of $18'' \times 18''$, where maximum oscillatory power was observed. This region was chosen to coincide with the sunspot umbra. We performed filtering of the sunspot oscillations during the observation period and found about 120 propagating wavefronts with an $\sim$3-minute periodicity. This number is sufficient to make preliminary conclusions about the direction of prevailing wave motion. In our study we selected two temporal intervals, 00:12:57 -- 00:14:57~UT and 00:48:09 -- 00:50:09~UT, during which wave activity increases and demonstrates characteristic wavefront shapes best defined as `spherical' and `linear'. Between these times we observed intermediate spatial shapes. Examining the set of cropped images revealed pronounced fluctuations in intensity along the bright features of the umbra. It can be assumed, and consistent with previous studies, that these details are associated with internal magnetic structures on which the waves propagate.

We have prepared a variance map (Fig.~\ref{2}e) of the active region to study the prevailing wave propagation paths. We found that the bright regions in the corresponding images coincide with peaks found in the variance map. There are two paths with variance increases in the Easterly and Westerly directions.  The broken arrows shows the prevailing sectors of wavefronts propagation in the Easterly and Westerly directions. The circle in Figure~\ref{2}(e) shows the localisation of the central pulsating source.

We applied the PMD technique to prepare the $\sim$3~minute broadband distribution of periods depending on their spectral power (Fig.~\ref{2}d) and a series of amplitude wavefronts (Fig.~\ref{3}) spanning 00:12:57 -- 00:14:57~UT and 00:48:09 -- 00:50:09~UT. To obtain the distribution of periods as a histogram, we searched for temporal differences (periods) between peaks present in all IMF components for all pixels in the image cube. The bin size was defined as a time resolution of the series. To obtain the reliability of distribution we prepared  their “smoothed” version with increased bin size. We see from Figure~\ref{2}d that the spectrum consists of a series of narrowband peaks (thin lines) covering the period range 2.0 -- 3.5~minutes. The fine structure of spectrum as width between peaks does not exceed $\sim$0.1~mHz, which is consistent with previously obtained results \cite{51}. The distribution with larger bin size (thick lines) highlighted the presence of multiple components. Maximum amplitude oscillations are localised with a period of $\sim$2.6~minutes. The power of the oscillations in the $\sim$3-minute window (Fig.~\ref{2}f) shows a thin spatial structure representative of elongated and localised sources inside the umbra. It can be assumed \cite{8} that each source is associated with a separate harmonic of the spectrum.

We found a compact pulsating source near the centre of the umbra, where spherical waves begin to propagate. This source is highlighted in Fig.~\ref{2}e and Fig.~\ref{3} (00:12:57 UT) as a circle, and is accompanied by the modulation of the $\sim$3-minute oscillations, visible as low-frequency wave trains with $\sim$10--20 minute periodicities \cite{52}. During the development of the wave trains, wavefront shape transformations occur, depending on the specific direction of motion. The movement paths are in good agreement with the direction of the bright regions visible in the variance map shown in Figure~\ref{2}e.

\begin{figure}[!t]
\vspace*{-7pt}
\centering\includegraphics[width=13cm]{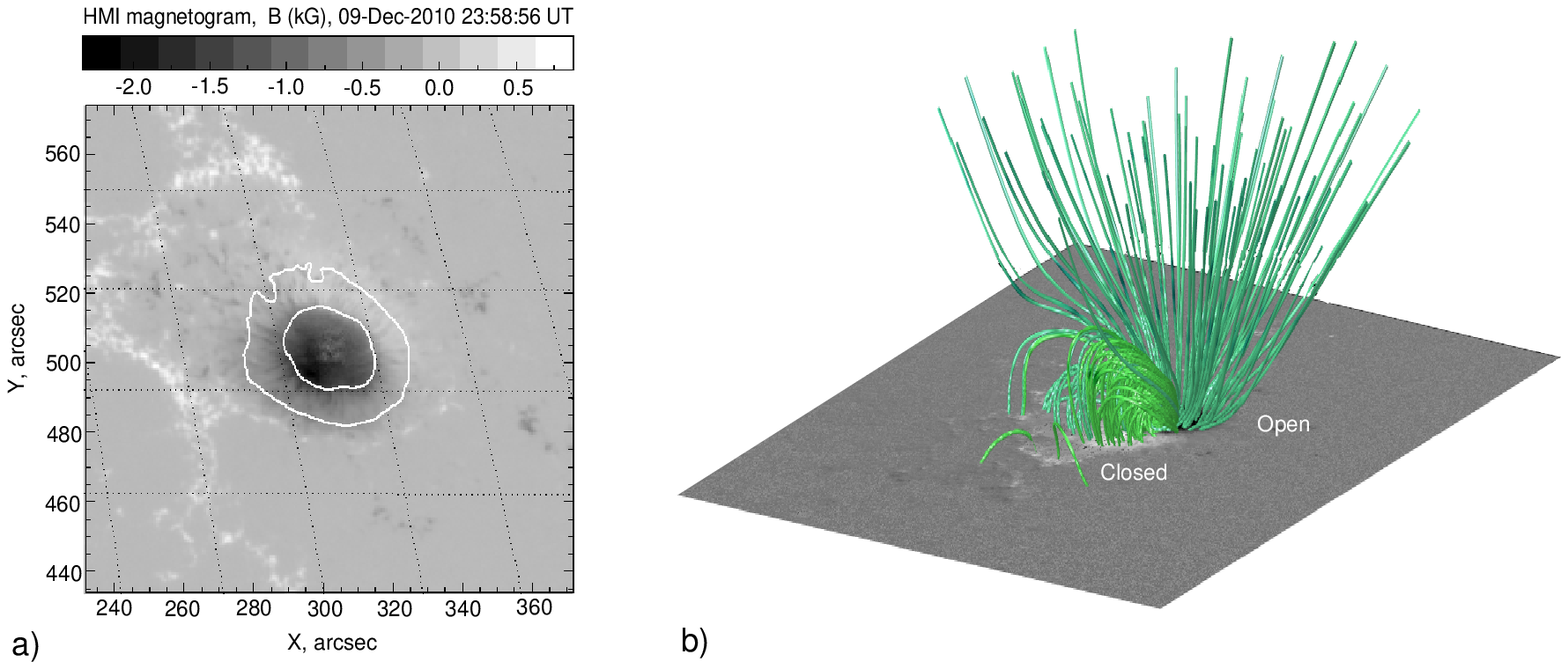}
\caption{(a) SDO/HMI magnetogram of active region NOAA 11131. (b) Three-dimensional non-linear force-free extrapolation of the magnetic field. The localisation of open and closed field lines is shown.} 
\label{4}
\end{figure}

We prepared the narrowband wavefronts visible during the observations using colour in Figure~\ref{3} to highlight the positive and negative amplitudes of the wavefront shapes. We see the evolving dynamics of the fronts in time through their motion in the spatial plane, alongside changing shapes and intensity oscillations. The arrows show the direction of the wave movements. Different colours are associated with the half-period parts of the wavefronts, where negative (yellow background and black contours) and positive (blue background and white contours) components are highlighted. The maximum brightness of the wavefronts was reached close to the umbral perimeter. We subsequently found a relationship between the direction of movement and the shape of wavefronts. For moving fronts in the Easterly direction, the quasi-spiral shape prevails (Fig.~\ref{3}, upper panels). Contrarily, when moving in the Westerly direction, the wavefronts transform into elongated linear shapes (Fig.~\ref{3}, bottom panels). The propagation of wavefronts from the centre to the umbral border only occurs in one direction (i.e., no return path towards the umbral core). For different wave trains, the individual directions may change, but not randomly, and only along the selected angular sectors. When moving, the central part of the fronts expand and become spherical due to the appearance of new arc-shaped components in the first case (upper panels of Fig.~{\ref{3}}) or the elongation of the linear fronts in the second case (lower panels of Fig.~{\ref{3}}).

To study the magnetic structuring of the sunspot active region, we employed vector magnetograms (Fig.~{\ref{4}}a) acquired by the Helioseismic and Magnetic Imager (HMI; \cite{2012SoPh..275..229S}) onboard the SDO spacecraft. The Very Fast Inversion of the Stokes Vector (VFISV; \cite{2011SoPh..273..267B}) algorithm was applied to SDO/HMI vector data acquired at the start of the 1~hour observing sequence. A $500\times500$~pixel$^{2}$ ($250\times250$~arcsec$^{2}$) region centred on NOAA~11131 was extracted from the full-disk magnetogram. The VFISV outputs are vector magnetic fields in a heliocentric coordinate system, which were converted into heliographic projections \cite{1990SoPh..126...21G }, providing magnetic field components parallel (Bx and By) and perpendicular (Bz) to the solar surface. Next, non-linear force-free field (NLFFF) extrapolations were performed \cite{2008JGRA..113.3S02W, 2016NatPh..12..179J}, with the resulting magnetic field line traces shown in the right panel of Figure~{\ref{4}}. The reliability of NLFFF extrapolations performed on highly magnetic sunspot structures has been documented in previous studies \cite{2013ApJ...779..168J, 2016ApJ...826...61A, 2018ApJ...860...28H, 2018NatPh..14..480G}, where the persistence of magnetically-dominated plasma (i.e., low plasma-$\beta$) with geometric height provides excellent suitability of NLFFF extrapolations in these locations. We see that anchored umbral field lines can be divided into `open' and `closed' categories. It can be assumed that the Eastern sector of waves propagation in Fig.~\ref{2}e are associated with low, closed magnetic loops connecting the lead sunspot to the trailing regions. The Western sector is associated with disturbances propagated along open, highly magnetic loop structures. Both of these structures are visible in the NLFFF extrapolations shown in the right panel of Figure~{\ref{4}}. This result coincides well with the magnetic structures observed in the SDO/AIA 171{\AA} coronal channel (Fig.~\ref{2}c) and the East-West asymmetry found earlier \cite{50}.

The derived dependence of the wavefront shape on both the direction of movement and the distance to the umbral core indicates the efficient filtering of broadband emission in both space and time. The main contribution is likely to be related to changes in the magnetoacoustic wave cutoff frequency. We can assume that the different inclinations of the magnetic field lines to the solar normal, alongside local spatial inhomogeneities, may lead to the appearance of the observed wave dynamics.

Previously, a number of publications showed a pronounced difference in the localisation of oscillation sources in sunspots \cite{53, 54}. In the umbra, sources with a period of $\sim$2 -- 4~minutes and maximum power are mostly localised and are not homogeneous across the entire umbra. For symmetrical sunspots, the sources are likely to take on a rounded shape. Lower frequency components, with periods up to $\sim$7~minutes, are located in the penumbra and demonstrate shapes consistent with nested rings. Their width and distance   from the umbral centre will increase as the period of oscillation increases. A similar distribution of narrowband sources was observed in umbra \cite{5}, with the main difference over penumbral structures being that they are located much closer to one another and their width only varies slightly with increasing distance from the umbral centre. This corresponds very well to the magnetic field model of a sunspot, viewed as an expanding bundle of thin magnetic tubes, vertical in the umbra and almost horizontal in the penumbra (Parker's model;  \cite{55}).

\begin{figure}[!t]
\vspace*{-7pt}
\centering\includegraphics[width=13cm]{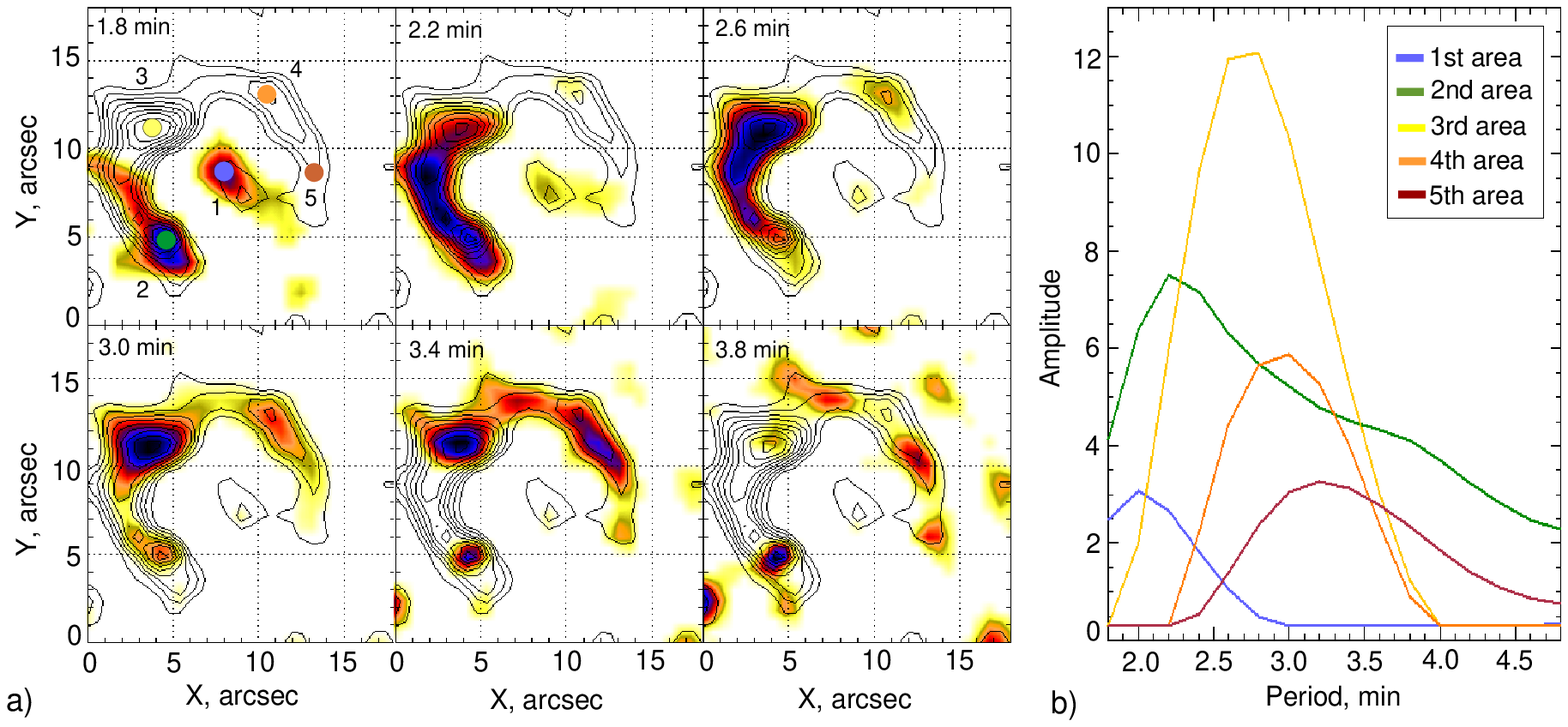}
\caption{The frequency distribution of oscillation power corresponding to the instantaneous wavefront visible at 00:14:33~UT (panel a). The contours show a broadband $\sim$3-minute wavefront superimposed on the narrowband power distributions corresponding to periods in the range of 1.8 -- 3.8~minutes. The corresponding power spectra for the selected points on the wavefront ridge, as a function of period. The coloured lines reflect the locations marked by the coloured dots in the upper-left panel. The first area is associated with a pulsating source.}
\label{5}
\end{figure}

For the studied highly-symmetrical sunspot, there is no clear spherical symmetry in the propagation of waves. However, there are clear changes to the wavefronts, both in time and in space, which can manifest as spherical, spiral, or linear transformations. During the presented observations, the direction of wave movement changes periodically, with wavefronts displaying a thin structure when bright sources of small angular size combine to create their macroscopic shape. This indicates that the simplified model of sunspot magnetic fields acting as a single expanding bundle of field lines does not correspond well to the results of our present observations. This may highlight the importance of studying the smaller-scale fine structures of umbral magnetic fields, which may help explain the results.

\subsection{Spatial and frequency structure of wavefronts}
To study the frequency structuring of the wavefronts, we applied the PWF technique \cite{33} to the SDO/AIA (304\AA) image cube, obtaining narrowband images with a positive half-wave period. In accordance with the power spectrum shown in Figure~\ref{2}d, we reconstructed the time signals for each pixel using a spectral filter consisting of a running narrowband window from $P_{i}/1.25$ to $P_{i}\times1.25$, in steps of 0.4~minutes, where $P_{i}$ is the current value of the period. The range of periods spanned 1.8 -- 3.8~minutes. For each reconstructed signal, the values of amplitude variations were calculated. 

\begin{figure}[!t]
\vspace*{-7pt}
\centering\includegraphics[width=13cm]{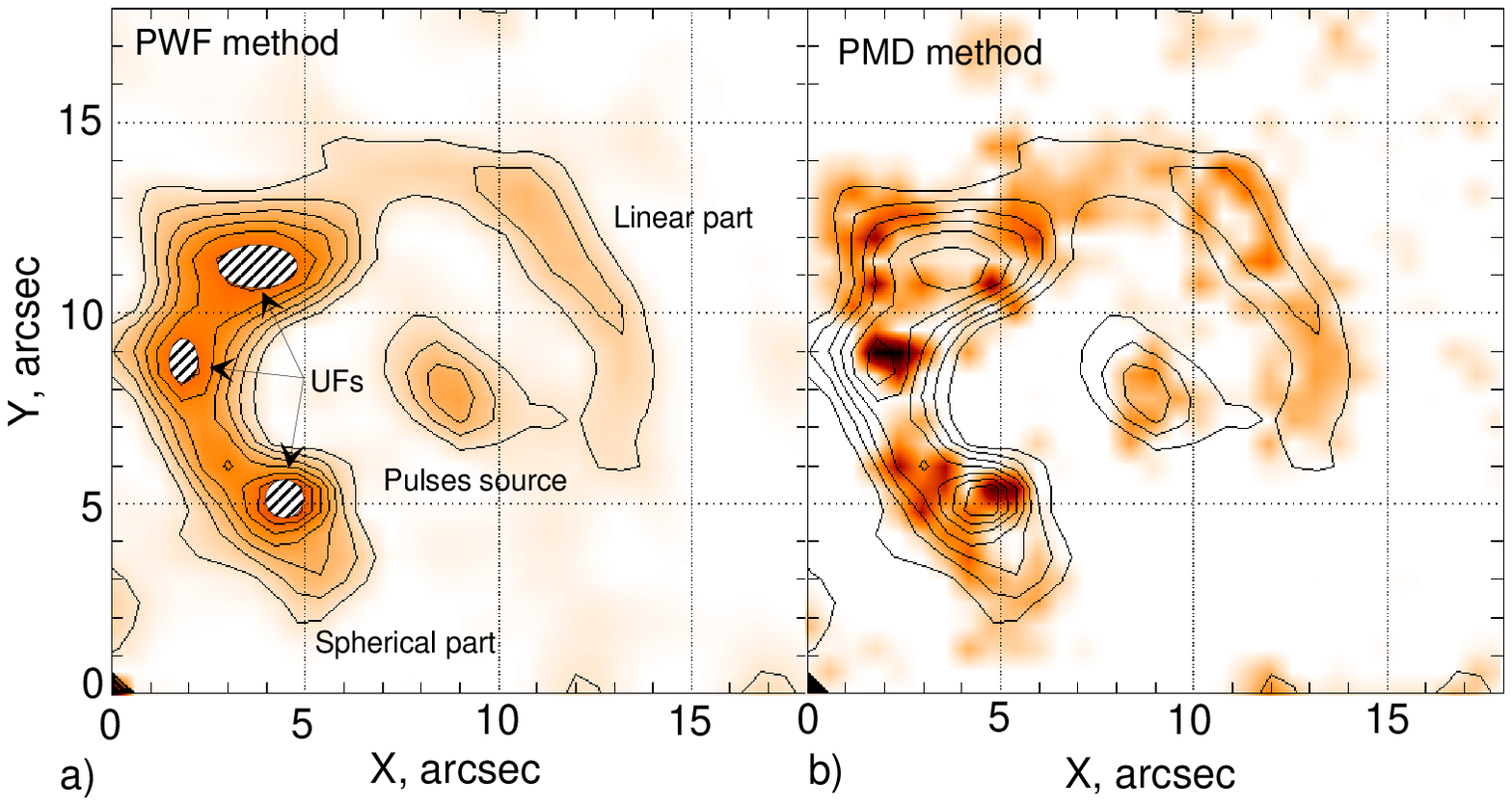}
\caption{Structure of the localised wavefront obtained using two different spectral techniques: (a) pixelised wavelet filtering and (b) pixelised mode decomposition. The filtering bandwidth is from 2 -- 4 minute periods, with the contours highlighting the broadband wave front. The bright parts of the wavefront as background umbral flashes are marked in the left panel and discussed in the main text.}
\label{6}
\end{figure}

Figure~\ref{5}a shows a snapshot obtained at 00:14:33~UT, where a quasi-spiral shape of the broadband wavefront as contours is clearly observed. Narrowband sources as a background are localised in different parts of the broadband wavefront, and their positions depend on the specific period they display. The high-frequency source with a period $\sim$2~minutes is located at the beginning of the wavefront. As the period increases, the central source marked with a blue circle in the upper-left panel of Figure~\ref{5}a disappears. There is a gradual increase in the brightness of individual parts of the wavefront, alongside a subsequent broadening of the maxima during its clockwise rotation. The peak brightness and, accordingly, the power of the oscillations reach their maximum values close to the central part of the wavefront, with a period of $\sim$2.6~minutes. Later, as the period grows, the brightness of the structure gradually reduces. We selected five control points, marked with coloured circles each representing the selected areas (Fig.~\ref{5}a), to trace the dependence of their wave power with oscillatory period. Each of the corresponding profiles shown in Figure~\ref{5}b has a peak value that shifts with the wave period. The maximum power is always located in the central part of the front, which is observed to move in the Easterly direction. 
		
For waves propagating in the Easterly direction along low, closed loops between 00:12:57 -- 00:14:57~UT (Fig.~\ref{3}), spherical or quasi-spiral wavefronts with a wide central structuring are observed. The characteristics are determined by the length of a narrowband wave path in the $\sim$3-minute range of periods. For low loops, this length will be the maximal. For waves that move in the opposite direction along high, open loops between 00:48:09 -- 00:50:09~UT, the wave path will be minimal. During this propagation, a linear spatial shape with a narrow central core will appear. We can conclude that the broadband front can be represented as the sum of individual narrowband sources. The observed spatial inhomogeneities of the wavefronts are determined by the contributions arising from each constituent spectral component. The appearance of a quasi-spiral shape depends on the direction and angle of the wave propagation. Different parts of the initially spherical front will radially propagate through the waveguides with different magnetoacoustic cutoff frequencies, which hence modifies their visibility in the $\sim$3-minute window. This will lead to a more complex spatial shape of the embedded wavefronts.

We applied the developed technique of PMD to highlight the fine spatial structure intrinsic to the $\sim$3-minute broadband wavefronts, and subsequently compared these results to those obtained by the PWF approach. To improve the signal-to-noise ratio and highlight the embedded sources, we summed over five half-cycle fronts at different times, corresponding to 00:12:09~UT, 00:13:21~UT, 00:14:33~UT, 00:15:45~UT, and 00:16:57~UT, whilst they moved through the same umbral location. Figure~\ref{6} shows the resulting images obtained using PMD and PWF techniques. 

\begin{figure}[!t]
\vspace*{-7pt}
\centering\includegraphics[width=7cm]{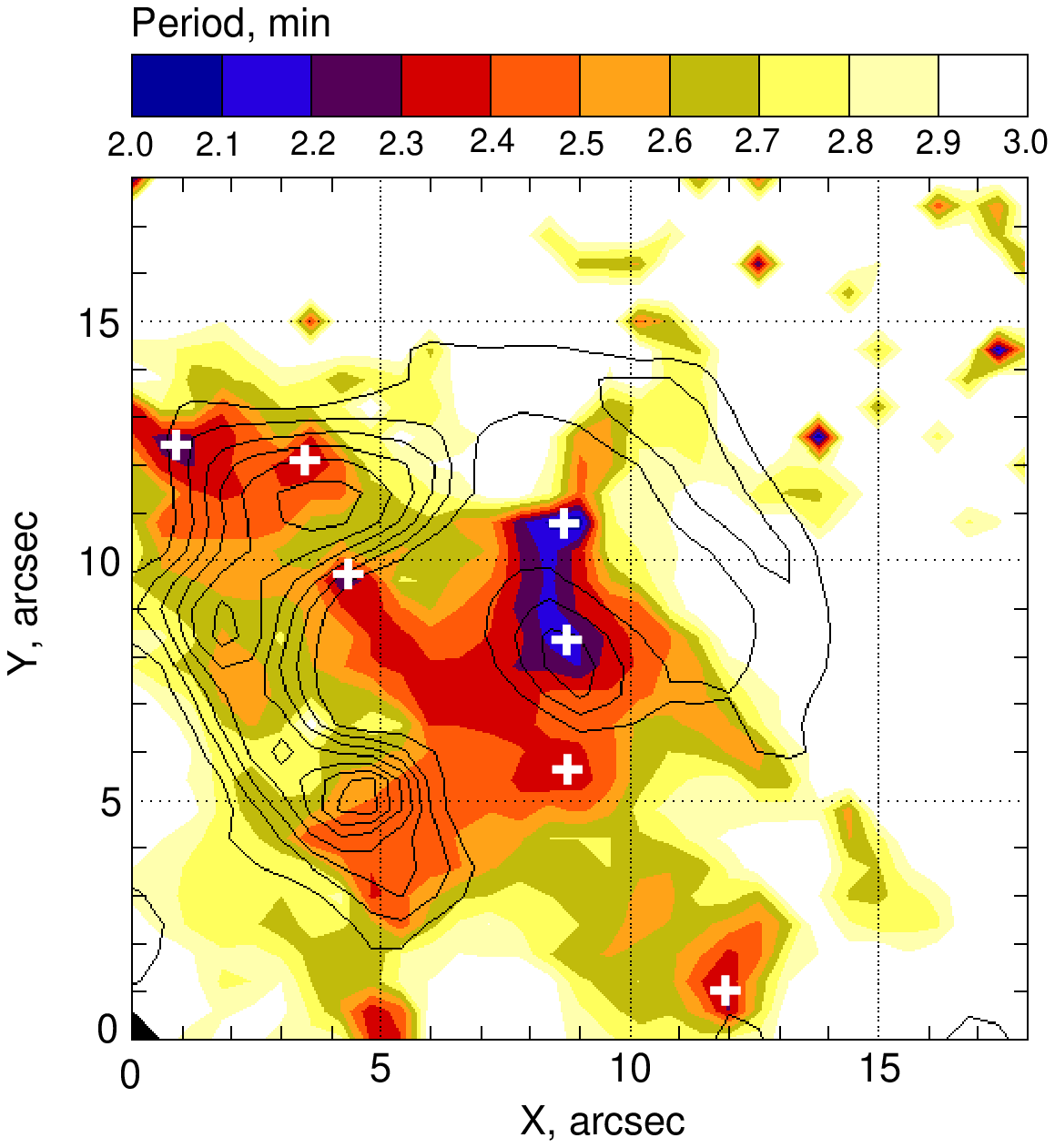}
\caption{A peak period map of the oscillations found within the sunspot umbra, where the colour bar indicates a different period in the range of 2.0 -- 3.0~minutes. The contours highlight the $\sim$3-minute broadband wave front at 00:14:33~UT, while white crosses indicate the location of compact oscillation sources discussed in the main text.}
\label{7}
\end{figure}
			
We see from Figure~\ref{6}a that the wavefront consists of a blurred quasi-spiral structure with five sources indicated by the numbers depicted in Figure~\ref{5}a. The spatial resolution of the PWF technique is not ideal when dealing with very small-scale wave sources. In the case of the PMD technique (Fig.~\ref{6}b), the wavefront is mapped by a set of clear stationary point sources with different brightnesses. These sources are located inside the contours obtained by the PWF technique. Some sources are grouped into oscillating clusters, with their angular sizes varying from $\sim$0.6 -- 3.0~arcsec. The brightest sources identified by the PWF and PMD techniques coincide with one another, but from Figure~{\ref{6}b} it is clear that the PMD approach allows very small-scale wave sources to be resolved with higher fidelity inside the umbra. 
		
The spatial location of the fine sources in Figure~\ref{6}b does not depend on the period. We observe only changes in their corresponding intensity. In addition, summing over five repeating wavefronts does not lead to blurring of these sources, showing persistent coherency in their signatures. Their appearance correlates well with background umbral flashes \cite{20} as locations of increased oscillations. These are placed on the wavefront background marked in Fig.~\ref{6}a as filled ellipses. We can conclude that these and the UF sources are the same spatial structures that are activated when the wave propagates through that spatial location. A sequence of narrowband frames revealed an increase in the brightness of the point sources when such radial perturbations moved through them. We found that as the period increases, the central part of the wavefront gradually disappears from view. Only symmetrical low-frequency sources remain at the front periphery. Narrowband sources, showing similar behaviours, were analysed in relation to the overall sunspot scale \cite{53}. Based on the dependence between the structure of narrowband images and the cutoff frequency of the waves, we can assume that we are observing a pattern linked to the embedded magnetic fields spanning the whole sunspot, but now on a much smaller spatial scale.

\subsection{Fine distribution of umbral magnetic fields}
Wavefronts as test sources have been used in the study of umbral inhomogeneities. Based on the relationships between the observed periods and the wave cutoff frequencies \cite{1}, we prepared maps of the magnetic field inclination angles and compared these with the localisation of visible wavefronts. The SDO/AIA 1600{\AA}, 304{\AA}, and 171{\AA} channels are used to obtain the dependence as a function of geometric height in the solar atmosphere. 

\begin{figure}[!t]
\vspace*{-7pt}
\centering\includegraphics[width=13.5cm]{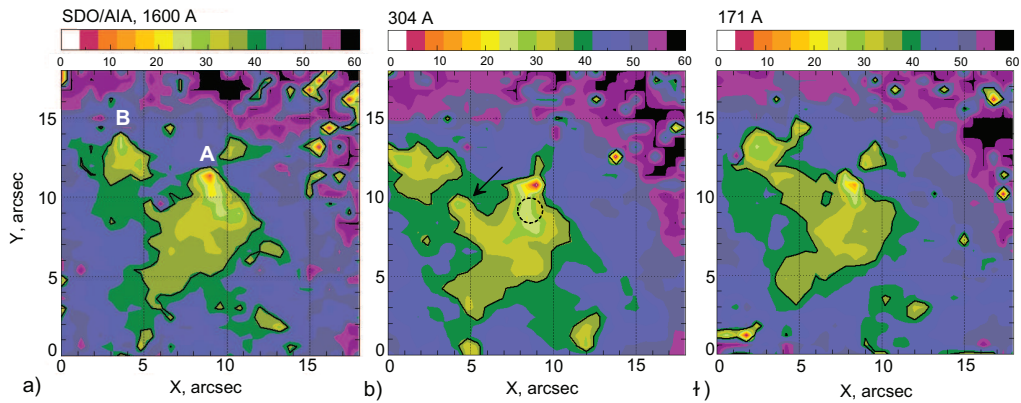}
\caption{The spatial distribution of magnetic field inclination angles with respect to the solar normal for different SDO/AIA channels, corresponding to the 1600{\AA} (left), 304{\AA} (middle), and 171{\AA} (right) filters. The region of interest discussed in the main text is highlighted using thick black contours. The letters `A' and `B' indicate the regions with predominantly vertical magnetic field lines. The arrow shows a bridge beginning to form between the two areas. The black circle shows the location of a pulsating source.}
\label{8}
\end{figure}

For each wavelength, we prepared colourised period maps based on the filtration of the images using the PMD technique, which was further validated using the global wavelet spectra calculated for all pixels using the PWF algorithms. The resulting power peaks at specific periods were mapped employing a strict 95\% confidence level. Figure~\ref{7} displays the peak period map obtained for the 304{\AA} channel, where the most visible wave activity was present. For visual clarity, we have superimposed a broadband wavefront as contours on top of the colour period map in Figure~{\ref{7}}.
		
From examination of Figure~{\ref{7}, we see that the distribution of periods is not symmetrical about the umbra. Inside the coloured ellipse, there are many different periods that cover parts of the umbra with the fine structuring synonymous with stable oscillation sources. In the central part there are two powerful high-frequency sources with an $\sim$2.3~minute periodicity at coordinates {[9$''$,~8$''$]} and {[9$''$,~11$''$]}. Their positions coincide with the location of the pulsating source defined in Figure~\ref{2}b using a circle. We also observed several similar knot-like sources with small angular sizes indicated by the white crosses in Figure~{\ref{7}}. Each of these knot-like sources can be connected via horizontal bridges with the same oscillation period, or they can remain on isolated islands separated from each other. All of the sources are surrounded by oscillations with increasing periodicities up to $\sim$3~minutes at the umbral boundary.


The observed distribution of periods in the umbra indicates the possible existence of stable horizontal magnetic structures with small angular sizes. Waves propagating along these features change their visibility depending on the cutoff frequency imposed by the inclination angle of the surrounding magnetic fields. Similar filamentary structures were discovered recently from observations in the Ca~{\sc{ii}} line during the development of UFs \cite{30}. Our results suggest that these structures are a common property of sunspot umbrae.

Based on the dominant period map shown in Figure~\ref{7}, we have computed a map of the field line inclinations with respect to the solar normal. For vertical field lines in sunspots embedded within a low plasma-$\beta$ environment (i.e., $\beta \ll 1$), the cutoff period depends on the local sound speed (i.e.,  temperature) and changes proportionally with the cosine of the angle of inclination \cite{1}. A larger inclination angle will allow low-frequency waves to penetrate upwards into the corona. Analysis of the influence of these values on the cutoff period showed that an increase in temperature at the sunspot periphery has a much smaller effect than the magnetic field inclination \cite{53}.  According to the empirical relation $P_{\text{cutoff}} (\phi) \approx 1.25~P_{\text{peak}} (\phi)$ \cite{56}, we can relate the peak period, $P_{\text{peak}}$, and the cutoff period, $P_{\text{cutoff}}$, to the field angle $\phi$. Based on the peak period for each location in Figure~\ref{7}, we reconstructed the angles of magnetic inclination using the simple formula $\phi (x, y) = \arccos \left(P_0 / P_{\text{cutoff}} (x, y)\right)$, where $P_0 \sim 2.7$~minutes is the local cutoff period for the active region. The calculated magnetic field inclination angles for the 1600{\AA} (upper photosphere), 304{\AA} (transition region), and 171{\AA} (corona) channels are shown in Figure~\ref{8}, allowing us track changes in the umbral magnetic fields with altitude. 

The inclination angles depicted in Figure~{\ref{8}} show slight changes across the different atmospheric heights. At the photosphere level (1600{\AA} panel in Fig.~\ref{8}a), there are two areas marked by bold black contours with sources designated as `A' and `B'. The  circle shows the location of a pulsating source. The inclination angles increase with radial distance from the sources, opening from $\sim$5~degrees through to $\sim$38~degrees. In the transition region (304{\AA} panel in Fig.~\ref{8}b), we observe the creation of an inclination angle `bridge', marked with an arrow in the middle panel of Figure~{\ref{8}}. In this location, there is a gradual steepening of the magnetic field lines. Significant changes in the angular slopes occur only in the Easterly direction. Other areas are more stable, demonstrating less changes. At the coronal level (171{\AA} panel in Fig.~\ref{8}c), regions `A' and `B' become merged, highlighting the connectivity between different layers of the solar atmosphere.

Based on obtained results, we can assume that the magnetic field of sunspots consist of not one large bundle of magnetic field lines, but from several bundles with different angular sizes and loop heights. Waves passing through these natural waveguides from sub-photospheric levels to the corona will have different oscillatory power and source sizes. We propose that there is a main bundle of magnetic field lines associated with the pulsating source highlighted in Figure~\ref{2}e. This shape is asymmetric, with some elongation in the spatial plane. The footpoint of this bundle is located near the high-frequency, central wave source labelled `A' in Figure~\ref{8}a, with the vertical field lines beginning to diverge with atmospheric height. We will observe an increase in the period of the oscillations depending on their distance from the centre of the umbra \cite{53}. We think that together with the main bundle of field lines, there is also a set of small magnetic bundles with footpoints, marked by white crosses in Fig.~\ref{7}. Spherically propagating waves passing through the umbral areas with different cutoff frequencies will acquire unique periods and oscillation power, forming instantaneous waveforms as seen in Fig.~\ref{6}. The different slope of the field lines leads to a change in the cutoff frequency and, accordingly, their narrowband visibility away from the footpoints. At different times we will see waves behaving differently. However, they all move along the same wave paths, which are formed by the underlying magnetic structures. The loops with large angles to the solar normal will be observed in the low photosphere. This means the chromospheric and coronal layers will be more representative of the steeper magnetic loops, leading to the appearance of a magnetic `bridge' structure as seen in Figure~\ref{8}c.

Magnetic inhomogeneities can be explained by both closure of a part of the field lines within the main magnetic bundle with neighbouring areas of opposite magnetic polarity, and by the closure of magnetic field lines between the small bundles positioned inside the umbral boundary. Their footpoints can be associated with the places demonstrating local changes in the oscillation period. A similar field structure with small-scale bipolar features was previously found in the penumbra, occurring between the more vertical fields of the penumbra and those associated with the horizontal fields that harbour the Evershed flow \cite{31,57}.

During propagating along low magnetic loops, the waves will change their visibility and shape due to the cutoff frequency. We observe high-frequency sources near the footpoints of local magnetic field bundles, with the displacement of low-frequency sources towards the periphery of the loops. Their spatial shape gradually changes from a pointed structure to a filamentary shape depending on the distance to the footpoint of the bundle. 

We assume that the sources found on the front ridge (see, e.g., Fig.~\ref{6}) are instantaneous snapshots of the parts of the magnetic waveguides hosting wave propagation. Their appearance is connected with background umbral flashes \cite{20} manifesting as brightenings in some parts of wavefronts. In these locations, the level of pulsing will be maximal. Applying the PWF spectral technique provides blurred narrowband sources, while the PMD technique allows us to isolate their small-scale signatures.
			
The observed magnetic structuring of the umbra can also explain the frequency drifts observed during the propagation of wave trains captured in radio and UV sequences \cite{15}. The beginning and end of these drifts coincide with the temporal dynamics of the trains. We can assume that when the wave propagates through the detected magnetic inhomogeneities with different gradients of the magnetic field inclination and, accordingly, the magnetoacoustic cutoff frequency, we will observe both positive and negative frequency drifts. When waves propagate along highly diverging loops, we will have a frequency change from high-to-low frequencies. For low loops, the reverse pattern of frequency drifts will be observed. Zero drifts are likely to be related to the propagation of wave trains through regions with the same magnetic field inclination. In the case of complete loop closure, multiple frequency changes will be observed within a single wave train traversing this type of structure.

\section{Conclusion}
We have developed a pixelised mode decomposition technique based upon empirical mode decomposition analysis. This allowed us to obtain amplitude, power, and phase characteristics of wavefronts in sunspots without using a pre-specified spectral filter. The technique was tested on an artificial datacube consisting of radially diverging waves with different periodicities and showed excellent suitability for solar studies. We compared the proposed PMD technique with the previously developed technique of pixelised wavelet filtering, revealing an increase in angular resolution when studying fine structures in oscillating sources. Using this technique, we were able to obtain a unique set of periodicities associated with the oscillation sources.

We found that wavefronts present during our observations propagated only along selected directions. These directions are associated with the orientation of the magnetic waveguides. During the development of the wavefronts, we observed that their shape changed depending on the direction of their movement. For travelling waves along low, fan-shaped closed loops that connect the lead sunspot to its trailing part, the wavefronts will have a broad spherical or quasi-spherical shape. On the other hand, when propagating along high, open loops, the wavefronts are transformed into linear shapes demonstrating a narrow central core. 
	
We used the PWF technique for spectral preparation of the SDO/AIA imaging cubes, and showed that broadband $\sim$3-minute wavefronts consisted of separate narrowband sources with different spatial shapes. These sources occur when waves pass through the inhomogeneities of the umbra. We assume that the different magnetic waveguides with different inclination angles and, accordingly, different cutoff frequencies, leads to the appearance of these sources. The simultaneous combination of narrowband spherical and linear sources leads to the appearance of visible spirality of the wavefronts in the $\sim$3-minute window.
	
Using the developed PMD techniques, we increased the spatial resolution and found structures inside the umbra whose widths neared the pixel platescale. These small sources remain stationary in space and change in brightness as increasing oscillations only when wavefronts pass through them. We interpret their appearance as the closure of low magnetic waveguides from the main (central) magnetic bundle to neighbouring regions with the opposite polarity. Derived maps of the magnetic field inclinations confirmed this assumption. We propose that the occurrence of these sources are associated with background umbral flashes. 

When examining the derived magnetic field inclination maps for different SDO/AIA channels, there is a dynamic evolution of the associated inclinations with atmospheric height for certain parts of the umbra. We observed only small variations in the magnetic field inclination for regions corresponding to the main magnetic bundle anchored in the umbral centre. Clear field line steepening was detected in locations away from the central umbral core, suggesting the presence of closed loop-like structures with small angular size.

We interpret previously detected frequency drifts in the $\sim$3-minute window to be a result of the fine inhomogeneities of the umbral magnetic field. Travelling waves propagating along different parts of the umbra with different gradients of the local magnetic field lines and, accordingly, differences in the local magnetoacoustic cutoff frequency, provide temporally resolved frequency drifts corresponding to the passage of the individual wave trains.

Using waves as test sources to track the magnetic structures of the sunspot atmosphere, we have shown that the umbra of NOAA~11131 has pronounced inhomogeneities. Wavefronts propagate radially along magnetic waveguides shown to possess different cutoff frequencies. This leads to their frequency fragmentation and the appearance of a quasi-spiral spatial shape. We have identified bundles of low magnetic loops that are stable in space and time. We propose that umbral flashes predominantly occur in these inhomogeneities during wave propagation. The development of next generation solar telescopes (e.g., DKIST, EST) with better spatial and spectral resolution will allow us to directly examine these fine-scale details in sunspots, understand their structure in both space and height, and validate our proposed hypotheses.


\dataccess{Data are available from the referenced sources or from the authors on request, and are courtesy of NASA/SDO and the AIA, EVE, and HMI science teams.}
\aucontribute{RS developed and tested the PWF and PMD spectral techniques. RS and DBJ calculated the peak period maps and derived the magnetic field inclination angles. DBJ calculated the three-dimensional extrapolation of the magnetic field. RS, DBJ and JS created the figures and wrote the paper.} 
\competing{The authors declare that they have no competing interests.}
\funding{The authors wish to acknowledge scientific discussions with the Waves in the Lower Solar Atmosphere team (WaLSA; www.walsa.team)  and the Royal Society through the award of funding to host the Theo Murphy Discussion Meeting "High resolution wave dynamics in the lower
solar atmosphere" (grant Hooke18b/SCTM). The study was performed within the basic funding from FR program II.16, RAS program KP19-270, and partially supported by the Russian Foundation for Basic Research (RFBR) under grant No. 17-52-80064 BRICS-a. RS research was funded by Chinese Academy of Sciences President’s International Fellowship Initiative, Grant No. 2020VMA0032. DBJ is funded by a Research \& Development Grant (059RDEN-1) provided by Invest NI and Randox Laboratories Ltd. JS research was partly supported by National Natural Science Foundation of China No. 11773038.}
\ack{The authors are grateful to the SDO/AIA/HMI teams for operating the instruments and performing the basic data reduction, and especially, for the open data policy. We are grateful to the referee for helpful and constructive comments and suggestions.}


\end{document}